\documentclass{PoS}

\usepackage{graphicx}
\usepackage{amssymb, amsmath}
\usepackage{subcaption}
\usepackage{rotating}
\usepackage{pdflscape}
\usepackage{xcolor}
\usepackage{ulem}

\usepackage{multirow}
\usepackage{longtable} 
\usepackage{mathrsfs}  
\usepackage{textcomp}
\usepackage{natbib}
\usepackage{graphicx}
\usepackage{subcaption}
\usepackage{natbib}
\usepackage{url}
\def\feii{Fe{\sc ii}}
\def\lledd{$L_\mathrm{bol}/L_\mathrm{Edd}$\/}
\def\rfe{$R_\mathrm{FeII}$}
\def\kms{km s$^{-1}$}
\title{Main trends of the quasar main sequence - effect of viewing angle}

\ShortTitle{Main trends of the quasar main sequence - effect of viewing angle}

\author{Swayamtrupta Panda\\
        Center for Theoretical Physics, Polish Academic of Sciences\\
        Nicolaus Copernicus Astronomical Center, Polish Academy of Sciences\\
        \href{panda@cft.edu.pl}{panda@cft.edu.pl}}
\author{Paola Marziani\\
        INAF-Astronomical Observatory of Padova, Vicolo dell'Osservatorio 5, 35122 Padova PD, Italy\\
        paola.marziani@inaf.it}
        
\author{Bo{\.z}ena Czerny\\
        Center for Theoretical Physics, Polish Academic of Sciences\\
        bcz@cft.edu.pl}

\abstract{We address the effect of viewing angle of the accretion disk plane and the geometry of the broad line region (BLR) with the goal of interpreting the  distribution of quasars along the main sequence (MS). We utilize photoionization code CLOUDY to model the BLR \feii{} emission, incorporating the grossly underestimated role of the form factor (\textit{f}).  We recover the dependence of  the strength of the \feii\ emission in the optical (\rfe{}) on \lledd{} ratio and related observational trends - as a function of the spectral energy distribution (SED) shape, cloud density, composition and intra-cloud dynamics, assumed following prior observations. With this approach, we are now able to explain the diversity of quasars and the change of the quasar properties along the Main Sequence (MS). Our approach also explains the rarity of the highest \feii\ emitters known as the extreme xA sources and can be used as a predictive tool in future reverberation mapping studies of Type-1 AGNs. This approach further justifies the use of quasars as `cosmological probes'.\\

Physical data and processes -- Accretion, accretion disks -- Line: formation -- Radiative transfer -- Turbulence -- Galaxies: active -- quasars: emission lines}

\FullConference{12th Serbian Conference on Spectral Line Shapes in Astrophysics Vrdnik, Serbia June 3-7, 2019\\
Contributions of the Astronomical Observatory Skalnate Pleso\\ 
ISSN (print): 1335-1842; ISSN (on-line version): 1336-0337 }
\begin{document}

\section{Introduction}
\label{intr}

From a theoretical scenario, a quasar spectrum can be modelled using 4 basic ingredients: (a) black hole mass; (b) mass accretion rate; (c) viewing angle; and (d) black hole spin (see \citet{campi18} for a recent review). \cite{bg92} incorporated the principal component analysis (PCA) to study the systematic trends between the numerous observed parameters of quasars. A long-standing issue in quasar astronomy has been the connection between observational and physical parameters \citep{50years}. The eigenvector 1 of the original PCA paved way for the quasar main sequence picture as we know it today \citep{sulenticetal00a, sh14}. The main sequence (MS) connects the velocity profile of `broad' H$\beta$ with the strength of the \feii\ emission (\rfe), i.e., the intensity of the \feii\ blend within 4434-4684 \AA\ normalized with the `broad' H$\beta$ intensity.

\feii\ is a complex ion that comprises of numerous multiplets and transitions. These transitions are produced via a number of line excitation processes (e.g. photoionisation, continuum fluorescence, collisional excitation, self-fluorescence within \feii\, fluorescent excitation by Ly$\alpha$ and Ly$\beta$ lines). The local physical conditions shape the spectrum and to make a deduction of these physical conditions, e.g., density, temperature, and iron abundance of the emitting regions, we require a complete simulation incorporating the various physical mechanisms that affect the \feii{} spectrum \citep{verner99}. In the new version of CLOUDY \citep{f13,f17}, the \feii{} emission is modelled with 371 levels up to 11.6 eV, including 68,535 transitions based on the \feii{} model of \citet{verner99}, which is a big improvement from the previous versions. The number of transitions is so large that the blended lines of \feii{} take the form of a pseudo-continuum, in which only few features can be unambiguously resolved. The modelled \feii{} pseudo-continuum shows quite good agreement with many observational \feii{} templates in the optical (\citealt{bg92,vc2003,kovacevic2010}).

In our previous works \citep{panda_frontiers, panda18b, panda19} we were successful in modelling almost the entire MS diagram constructed for over 20,000 SDSS quasars. In the past studies, we had incorporated only two of the aforementioned physical parameters of the super massive black hole i.e., black hole mass and accretion rate. The modelling was affected also by the cloud density, metallicity and microturbulence.

In \citet{panda19b}, we have shown that taking into account the viewing angle along with systematic trends in \lledd{}, local cloud density, cloud chemical composition and the shape of the ionizing continua which are known from prior observations, we can (a) explain the quasar main sequence starting from the low-\rfe\ high FWHM sources (Population B, FWHM H$\beta$ $> $ 4000 \kms) to the high-\rfe\ low FWHM sources (Population A, FWHM H$\beta$ $\le $ 4000 \kms); and (b) explain why the highly accreting sources are also high-\feii\ emitters. These sources are of special importance in view of their potential use  as  Eddington standard candles i.e., sources for which the \lledd{} and not the luminosity can be assumed to scatter around a well-defined value \citep{wangetal13, ms14}.

In the current work, we present the model which incorporates the viewing angle effects in a much more careful way. Instead of discreet values of the density and metallicity, fixed for each AGN spectral type along the main sequence, we now allow for a range of densities and metallicities in each bin. We also analyze the effect of the SED, microturbulence and the black hole mass. The method allows to obtain more generic constraints for a viewing angle for each spectral type class. However, the method is now much more computational time consuming so at present we show the full results only for one representative spectral bin A1. In Section \ref{methods}, we describe the derivation of the form factor and how it is incorporated in our modelling. In Section \ref{results}, we describe the outcomes from these photoionisation simulations performed with \textit{CLOUDY} in terms of (a) viewing angle, (b) shape of the SEDs, (c) micro-turbulence, and (d) increasing black hole mass. We show the relevance of this model to be used as a predictive tool to estimate the BLR sizes. In Section \ref{conclusions}, we summarize the results and provide a road-map for the future work.

\section{Method and Analysis}
\label{methods}

We perform the theoretical  modelling of the quasar properties assuming, as in \citet{panda_frontiers, panda18b}, that the central black hole is surrounded by an accretion disk  which provides the optical/UV continuum, and a hot corona, which is the source of the X-ray radiation. This continuum illuminates the BLR clouds located at a  distance given by the BLR size. Photoionization modeling allow us to calculate BLR line intensities, and the BLR radius is reflected in the kinematic line width under the assumption of the Keplerian motion.

However, there are several new elements in the current study. First, we now allow for the dependence of the AGN appearance on the viewing angle, instead of using universal (average) viewing angle for all objects as in \citet{panda18b,panda19}. Second, we approximate the trends noticed before between the model parameters and the source location in the \rfe{} - FWHM(H$\beta$) plane, and we model separately the spectral bins of the quasar MS plane assuming the representative values appropriate for each bin. Below we describe step by step our new approach to quasar MS modelling.
\subsection{Effect of viewing angle on the main sequence}
\label{viewing-angle}
The virial relation is used to estimate the inner radius ($r_{BLR}$) of the broad, ionized cloud \citep{woltjer1959}:
\begin{equation}
    r_{BLR} \propto \frac{G M_{BH}}{\sigma_{line}^2}
    \label{eq1}
\end{equation}
where, $G$ is the Gravitational constant, $M_{BH}$ is the mass of the black hole, and, $\sigma_{line}^2$ is the square of the velocity dispersion of the emission line that is considered. This velocity dispersion can be replaced with the line's full-width at half maximum (FWHM) which is the radial velocity projection of the the "\textit{true}" Keplerian velocity ($v_\mathrm{k}$). Replacing the proportionality sign with a constant, we can write
\begin{equation}
    r_{BLR} = \left(\frac{1}{f}\right)\frac{G M_{BH}}{FWHM^2}
    \label{eq2}
\end{equation}
The line FWHM can be expressed as:
\begin{equation}
    FWHM^2 = 4\left(v_\mathrm{iso}^2 + v_\mathrm{K}^2 \sin^2\theta \right)
    \label{eq3}
\end{equation}
where $v_\mathrm{iso}$ is the isotropic velocity component and $\theta$ is the viewing angle \citep{collin2006}. The viewing angle is defined as the angle between the axis perpendicular to the disc and the line of sight to the observer. The full range of the viewing angles considered for the modelling is 0-60 degrees. This range is chosen to select only those sources that are un-obscured in accordance with unification schemes (\citet{antonucci93,urrypadovani95}, see \citet{padovani2017} for a recent review). The FWHM is related to the $v_\mathrm{k}$ by
\begin{equation}
    v_\mathrm{k}^2 = f FWHM^2
    \label{eq4}
\end{equation}
\par 
This proportionality constant has been a factor of debate (see \citealt{Yu2019} for a recent review and references therein). Known as form factor (or structure factor or virial factor), $f$ depends on the structure, kinematics, and inclination of the BLR \citep{collin2006}. Combining Equations \ref{eq3} and \ref{eq4}, we get
\begin{equation}
    f = \frac{1}{4}\left[\frac{1}{\kappa^2 + \sin^2\theta}\right]
\end{equation}
where, $\kappa$ is the ratio between $v_\mathrm{iso}$ and $v_\mathrm{K}$, which decides how isotropic the gas distribution is around the central potential. If the value is close to zero, it represents a flat disk with thickness almost zero. On the other hand, if the value of $\kappa$ is close to unity, it represents an almost spherical distribution of the gas.  
\par

\begin{figure}
    \centering
    \includegraphics[width=\textwidth]{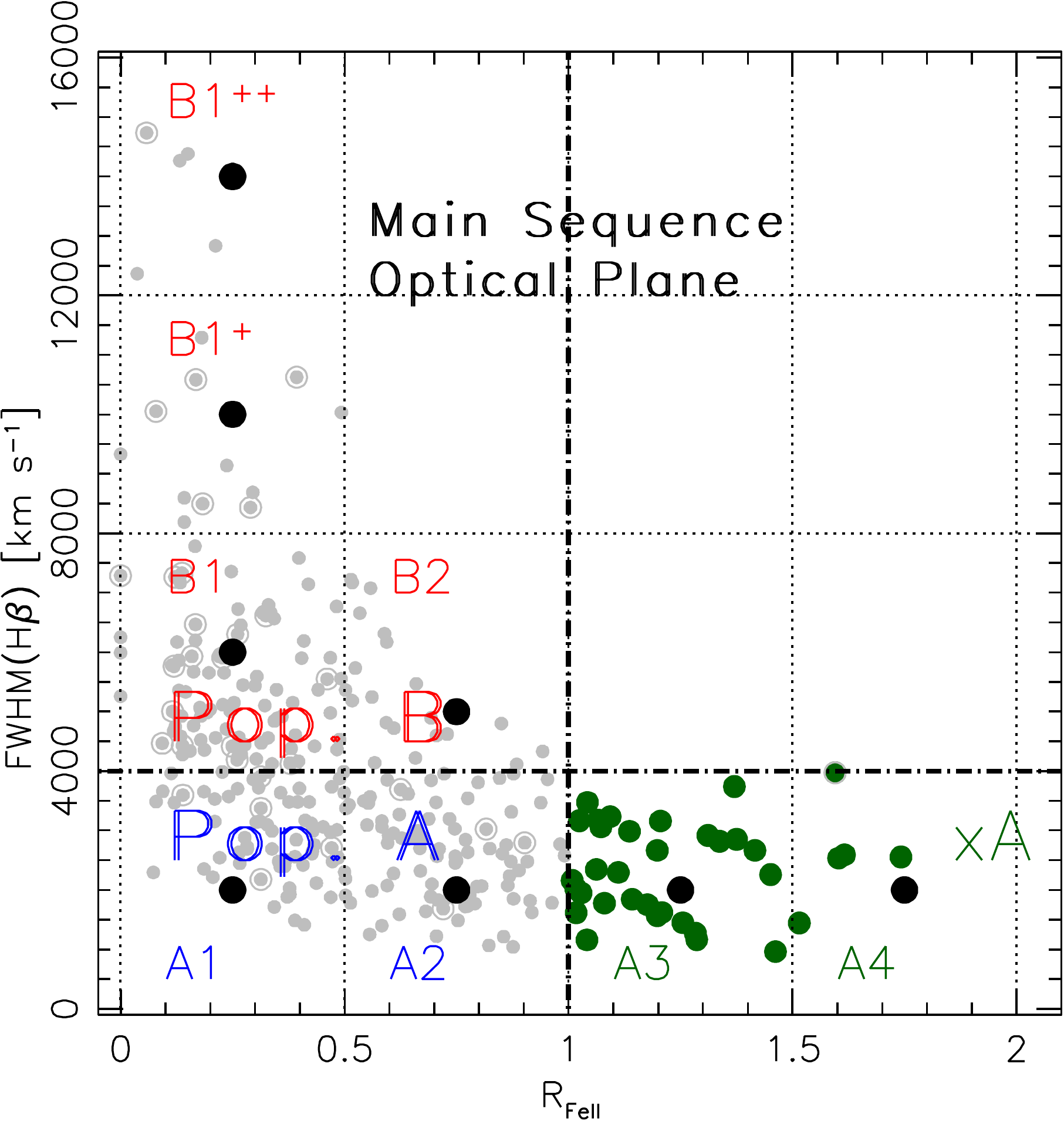}
        \caption{The diagram shows the optical plane of the Eigenvector 1 MS, FWHM(H$\beta$) vs. \rfe{}. The shaded area indicatively traces the distribution of a quasar sample from \citet{zamfiretal10}, defining the quasar main sequence. Circled symbols identify radio-loud sources. The thick horizontal dot-dashed line separates populations A and B. The plane has been further subdivided in spectral bins as defined by \citet{sulenticetal02}. The vertical dot-dashed line marks the limit for extreme Population A (xA) sources with \rfe\ $\gtrsim$ 1, whose data-points are in dark-green colors. The large black  dots mark the average values in each spectral type which are incorporated in the modelling. The diagram is adapted from  Fig. 2 of  \citet{marziani_frass18}.}
    \label{fig:inset}
\end{figure}


\subsection{Photoionisation modelling of the main sequence}

We use the latest version of the publicly available photoionisation code \textit{CLOUDY} \citep{f17} to solve the radiative transfer under local thermodynamic equilibrium (LTE),  satisfying ionization balance under a plane-parallel approximation. We assume a single cloud model where the density ($n_\mathrm{H}$) of the ionized gas cloud is varied from $10^{9}\; \mathrm{cm^{-3}}$ to $10^{13}\; \mathrm{cm^{-3}}$ with a step-size of 0.25 (in log-scale). We utilize the \textit{GASS10} model \citep{gass10} to recover the solar-like abundances and vary the metallicity within the gas cloud, going from a sub-solar type (0.1 Z$_{\odot}$) to super-solar (100 Z$_{\odot}$) with a step-size of 0.25 (in log-scale). The size of the BLR is estimated from the virial relation, assuming a black hole mass, a distribution in the viewing angle [0-90 degrees] and FWHM as described in Section \ref{viewing-angle}. The total luminosity of the ionizing continuum is derived assuming a value of the \lledd{} and the respective value for the black hole mass. We utilize different SEDs to highlight the differences in the shape of the ionizing continuum that is found to be relevant for the recovered values of the intensities of the line emissions (see \citet{panda19b} for more details).

 
\section{Results and Discussions}
\label{results}

\subsection{Interpretation of the quasar main sequence}

As an initial test, we assume a fixed black hole mass ($M_{BH} = 10^8 M_{\odot}$). Two Populations are identified in Figure \ref{fig:inset} -- the quasar main sequence diagram: Population A with FWHM H$\beta  \le $ 4000  \kms, and Population B of sources with H$\beta$\ broader than 4000 \kms\ (for a rational justification about the distinction between the two populations and a description of the main systematic differences, see \citealt{marziani_frass18,marzianietal19}). We consider the mean FWHM values in each spectral type depicted in Figure \ref{fig:inset} to see the effect of changing the FWHM across the main sequence plane i.e., going from the lowest FWHM to the extreme Population B sources that have the highest values of FWHM $\gtrsim$ 10000 km/s. Hence, as a first test, we take the mean values in each spectral bin For the Population A spectral types, we use FWHM = 2000 \kms; for Population B spectral type B1, FWHM = 6000 \kms\ and for B1$^+$ FWHM = 10000 \kms. We also investigate the B2 spectral type and use FWHM = 5000 \kms\ in this case. This is justified as the source count in this spectral bin (B2) is quite low ($\sim$3\%) and spans out the lower triangular region. The rationale to separate two quasars populations stems from the spectral differences recognizable by eye: Pop. A sources usually show low [OIII] emission, the strongest FeII emission; Pop. B sources give the impression of a much higher degree of ionization, with weak FeII, prominent [OIII] and CIV$\lambda$1549 emission. By considering trends in metallicity, density, and Eddington ratio derived from earlier work, we were able to account for the \rfe\ values (even for the highest ones!) in the spectral bins of the MS, and gain constraint on the viewing angle assuming an appropriate value of the BLR radius. These results are presented and discussed in \citet{panda19b}. For the purpose of this paper, we will only concentrate on a representative case i.e., spectral type A1. 

\subsection{Constraints on the viewing angle}
\label{viewing-angle}
The use of angle-dependent form factor (see Eq. 2 and 5) is a crucial extension of the quasar MS modelling done by \citet{panda_frontiers, panda18b, panda19}. Viewing angle directly affects the line width, and the object luminosity, and indirectly the estimated distance to the BLR and the location of an object on the quasar MS diagram. Inverting this dependence, we can obtain interesting constraints on the viewing angle of a source if its the location on the quasar MS diagram and the BLR size are known.

\begin{figure}
\hfill
\vfill
\begin{turn}{90}
\begin{minipage}[c][1\textwidth][c]{1\textheight}
\begin{subfigure}[h]{.475\linewidth}
\includegraphics[width=\linewidth]{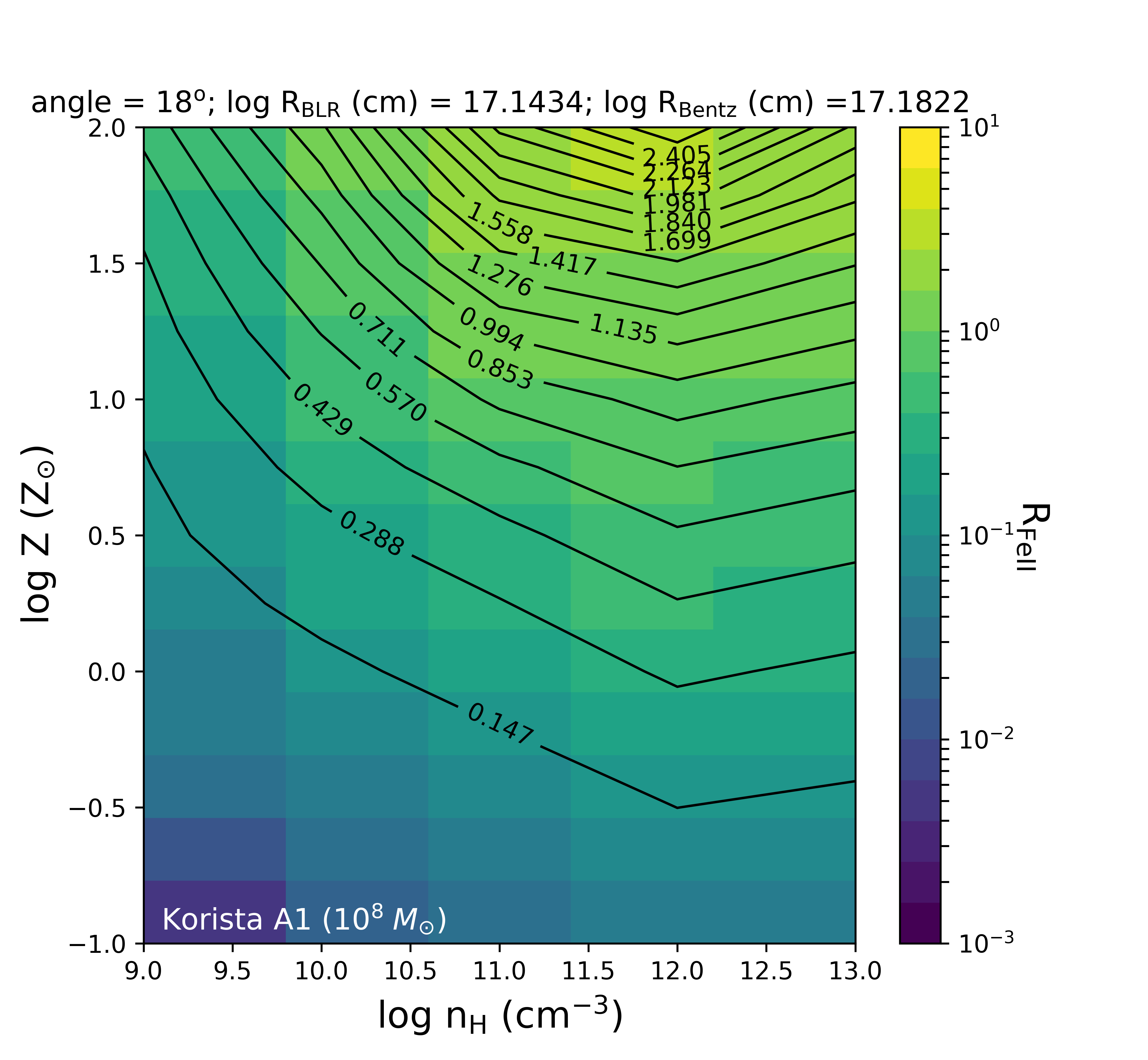}
\end{subfigure}\hfill
\begin{subfigure}[h]{.475\linewidth}
\includegraphics[width=\linewidth]{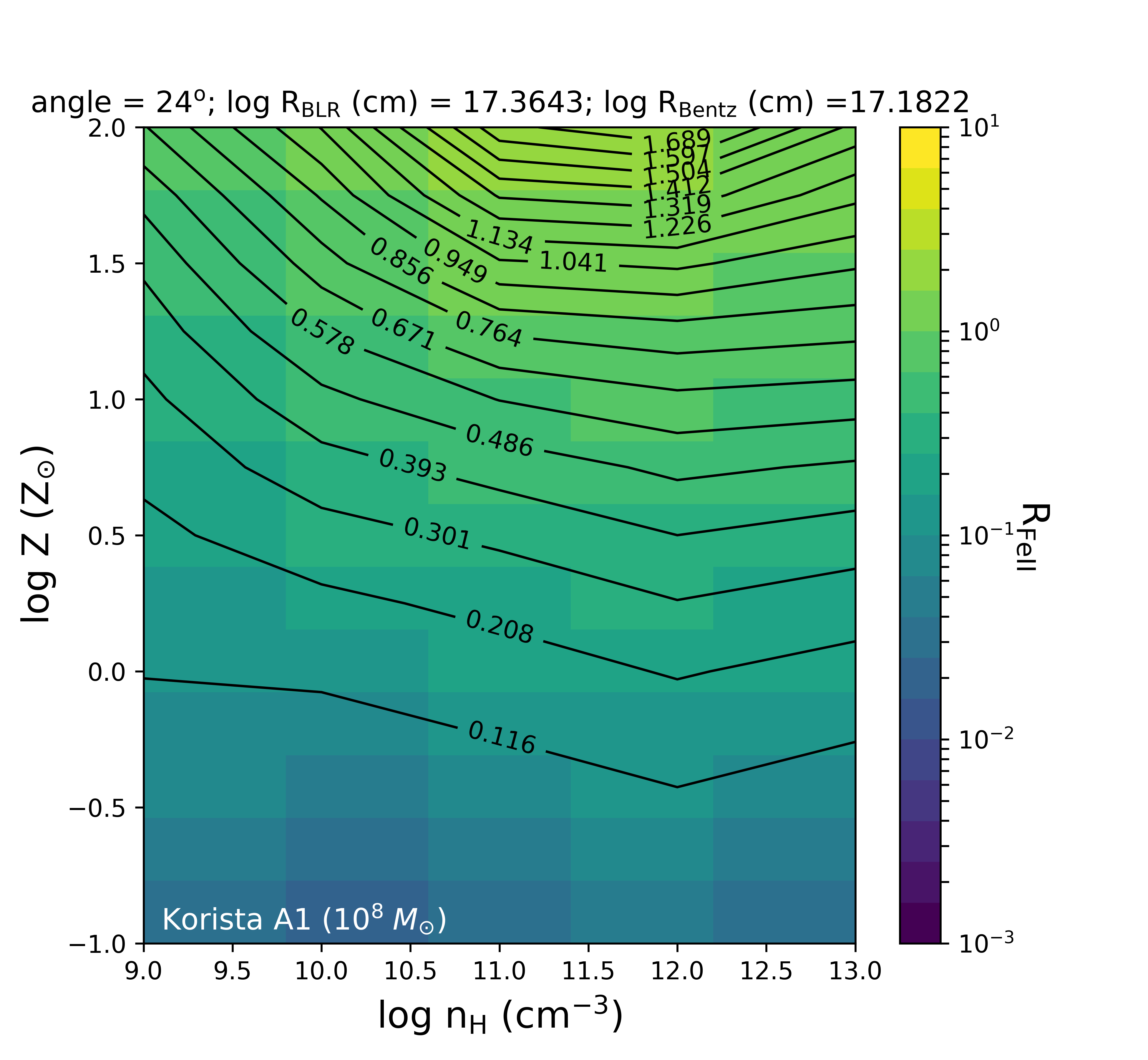}
\end{subfigure}
\caption{Constraints on the viewing angle. The figure shows two 2D density plots which map the distribution of the cloud density as a function of the metallicity. The colorbar depicts the value of \rfe\ . The plots show the results from a set of CLOUDY simulations for consecutive cases in viewing angle (18 degrees and 24 degrees). Parameters shown correspond to $M_\mathrm{BH} = 10^8\; M_{\odot}$, at zero turbulence, and using an SED shape taken from \citet{kor97}. The plots shown are for a representative case of the spectral type A1 where the mean of the FWHM is assumed at 2000 \kms\  with an assumed Eddington ratio, $\lambda_{\rm{Edd}}$ = 0.2. The corresponding values of the $r_\mathrm{BLR}$ computed from Equation \ref{eq2} and the respective $r_\mathrm{BLR}$ from the standard $r_\mathrm{BLR}$-$L_{\mathrm{5100}}$ relation are shown in the title of each plot.  \href{https://drive.google.com/file/d/1YBb9zHYNWgxVzCVZu99qUBxh6AEbjVHl/view?usp=sharing}{(link to the animation)}}
\label{fig:montage_m8_Kor_A1}
\end{minipage}
\end{turn}
\end{figure}

In Figure \ref{fig:montage_m8_Kor_A1}, we show two consecutive snapshots from one of the simulation results ($\theta = 18^{\rm{o}}$ and $\theta = 24^{\rm{o}}$). The plots show the distribution of \rfe\ as a function of gas density ($n_\mathrm{H}$) and  cloud composition (in terms of the metallicity, $Z$). The colorbar represents the values of \rfe\ which are also reported in the form of overlaid contours of the 2D distribution of $Z$ and $n_\mathrm{H}$ as a function of \rfe{}. The case shown is for a black hole mass, $M_\mathrm{BH} = 10^8\; M_{\odot}$ and for a bolometric luminosity, $L_\mathrm{bol}$ = 0.2$L_\mathrm{Edd}$. To obtain the monochromatic luminosity at 5100\AA, we utilize the mass and the Eddington ratio and compute the bolometric luminosity. Using the normalization coefficient as a function of mass and accretion rate, we estimate the monochromatic luminosity at 5100\AA\ (see Equation 5 in \citet{panda18b} for details).

The simulations were performed with a step size of 6 degrees in the viewing angles. Yet, they already show the tight constraints in the viewing angle that can be inferred for a fixed black hole mass. The viewing angle is correspondingly connected to the  radius of the BLR ($r_\mathrm{BLR}$) that is derived from the virial relation. In the simulation arrays of Fig. \ref{fig:montage_m8_Kor_A1}, the $r_\mathrm{BLR}$\ from the virial relation imposed to be close to the one  predicted from the standard $r_\mathrm{BLR}$-$L_{\mathrm{5100}}$ relation \citep{bentz13}\footnote{The viewing angle, $r_\mathrm{BLR}$ from the virial relation and the corresponding $r_\mathrm{BLR}$ from the \citet{bentz13} relation are reported in the title of each plot in the Figures \ref{fig:montage_m8_Kor_A1}, \ref{fig:seds}, \ref{fig:turb} and \ref{fig:montage_m8_m10}.}. 


Figure \ref{fig:montage_m8_Kor_A1} only shows a representative case with SED shape taken from \citet{kor97} and for a representative spectral type A1 (FWHM = 2000 \kms). In \citet{panda19b}, we confined the physical parameters in each spectral bin (see Table 1 in \citealt{panda19b}), for example, in A1, the cloud density ($n_{H}$) was assumed to be $10^{10.5}$ cm$^{-3}$, the metallicity was fixed at 5Z$_{\odot}$ and the bolometric luminosity was 0.2L$_{\rm{Edd}}$. In this paper, we remove these restrictions and consider the full parameter range with respect to density, metallicity and Eddington ratio.

From Figure \ref{fig:inset}, we can obtain the range of the \rfe{} for the spectral type A1 -- [0,0.5]. Taking this upper limit and comparing it with the panels in Figure \ref{fig:montage_m8_Kor_A1}, we see that contrary to our previous assumption -- the local cloud density is an almost constant entity within each spectral type, the density actually shows a broad distribution. In the considered grid on local densities i.e., from $10^{9}\; \mathrm{cm^{-3}}$ to $10^{13}\; \mathrm{cm^{-3}}$, the corresponding value of the \rfe{} ([0,0.5]) can be recovered. But we find that there exists a coupling between density and metallicity that was not studied in the previous works, that is, relatively higher densities require slightly super-solar metallicities ($\sim$3-5Z$_{\odot}$), while for the lower density cases the metallicity can be as high as $\sim$100Z$_{\odot}$. This suggests that the BLR cloud can indeed exist at densities higher than those predicted from radiation-pressure confinement estimates, i.e. $\sim$10$^{11}\; \mathrm{cm^{-3}}$ \citep{bl14}. With an increase in the viewing angle under the same parameterization, we see that there is a requirement of higher metallicities, especially for the higher densities ($> 10^{10}\; \mathrm{cm^{-3}}$) which goes up by a factor $\sim$2.

We have made an extensive set of simulations with a broad range of parameters and we will discuss this in detail in a forthcoming paper (Panda et al. in prep).

\subsection{Comparison of the spectral energy distributions}

Motivated by the fact that there is a broad distribution of quasars in the main sequence and that no singular ionizing continuum shape can effectively explain all the quasars, we have used 4 different spectral energy distributions that are appropriate for explaining the sources based on their spectral types. In \citet{panda19b}, each spectral type in the MS diagram, namely in Population A (A1, A2, A3, A4 and A1$^{*}$)  and in Population B (B1, B1+ and B2) is modelled with a specific SED shape. In this paper, our approach is different and we incorporate the four SEDs for spectral type A1 to understand the effect of the shape of the ionizing continuum which affects the recovery of the \rfe{}. These SEDs are adopted from \citet{mf87}, \citet{kor97}, \citet{laor97} and \citet{ms14}. 
Figure 1 in \citet{panda19b} shows the differences in these SED shapes, especially in the 1 - 25 Rydberg energy range which corresponds to the the optical-UV bump feature in a characteristic quasar SED i.e. the Big Blue Bump \citep{czerny87, richards06}.
\par
\begin{figure}[b]
\centering
\begin{subfigure}{.5\textwidth}
    \centering
    \includegraphics[width=1.075\textwidth]{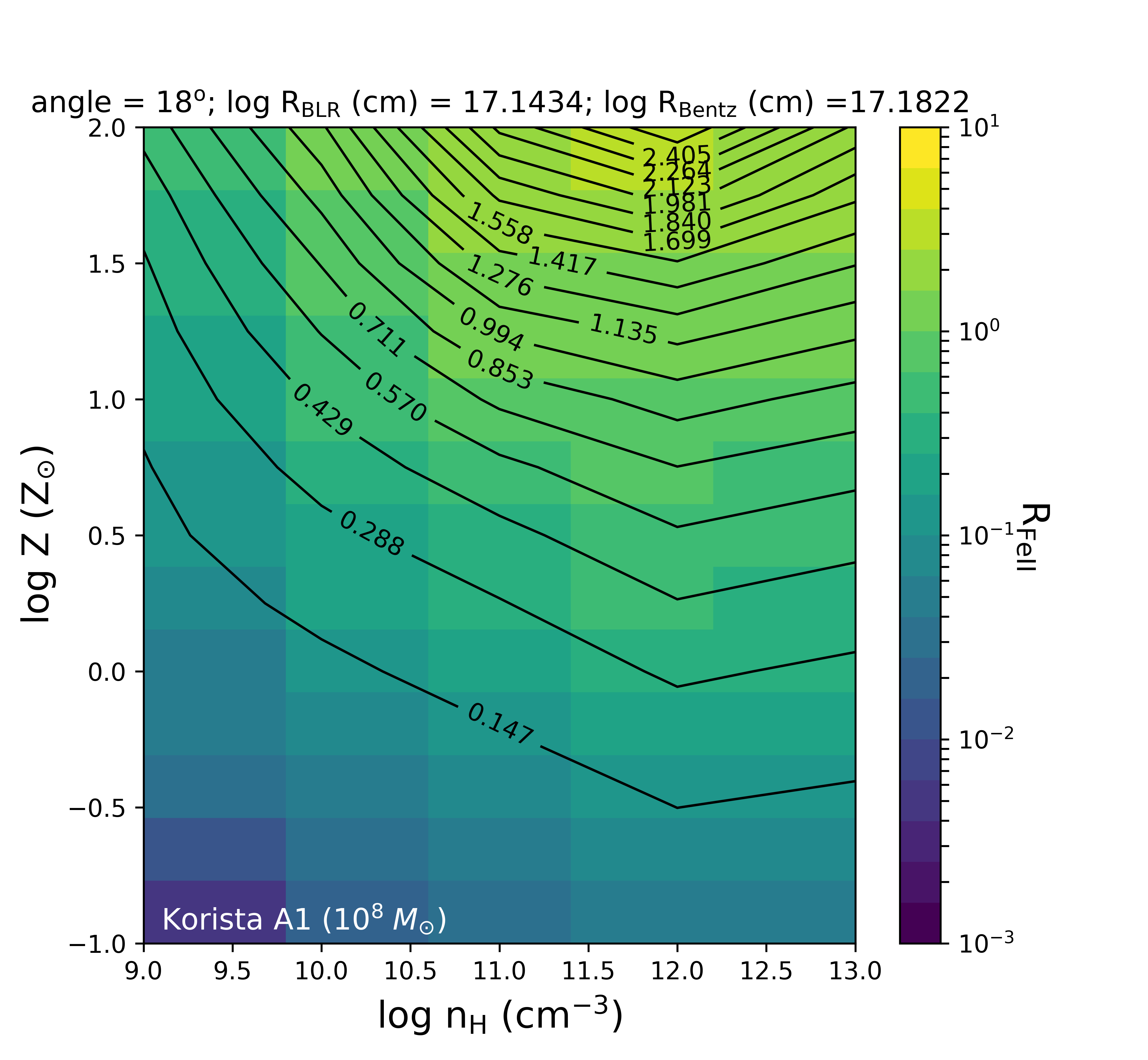}
\end{subfigure}%
\begin{subfigure}{.5\textwidth}
    \centering
    \includegraphics[width=1.075\textwidth]{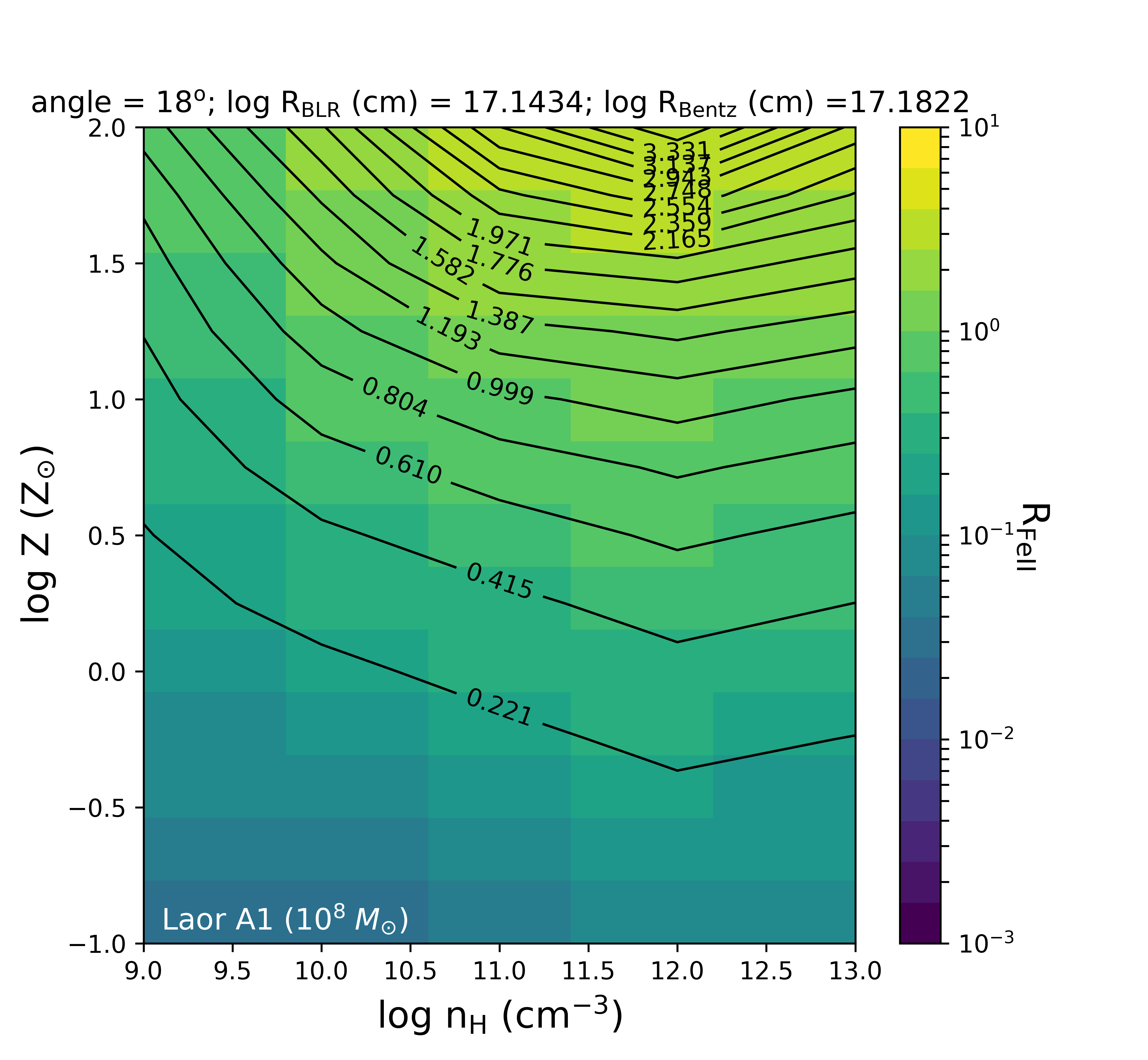}
\end{subfigure}
\vfill
\begin{subfigure}{.5\textwidth}
    \centering
    \includegraphics[width=1.075\textwidth]{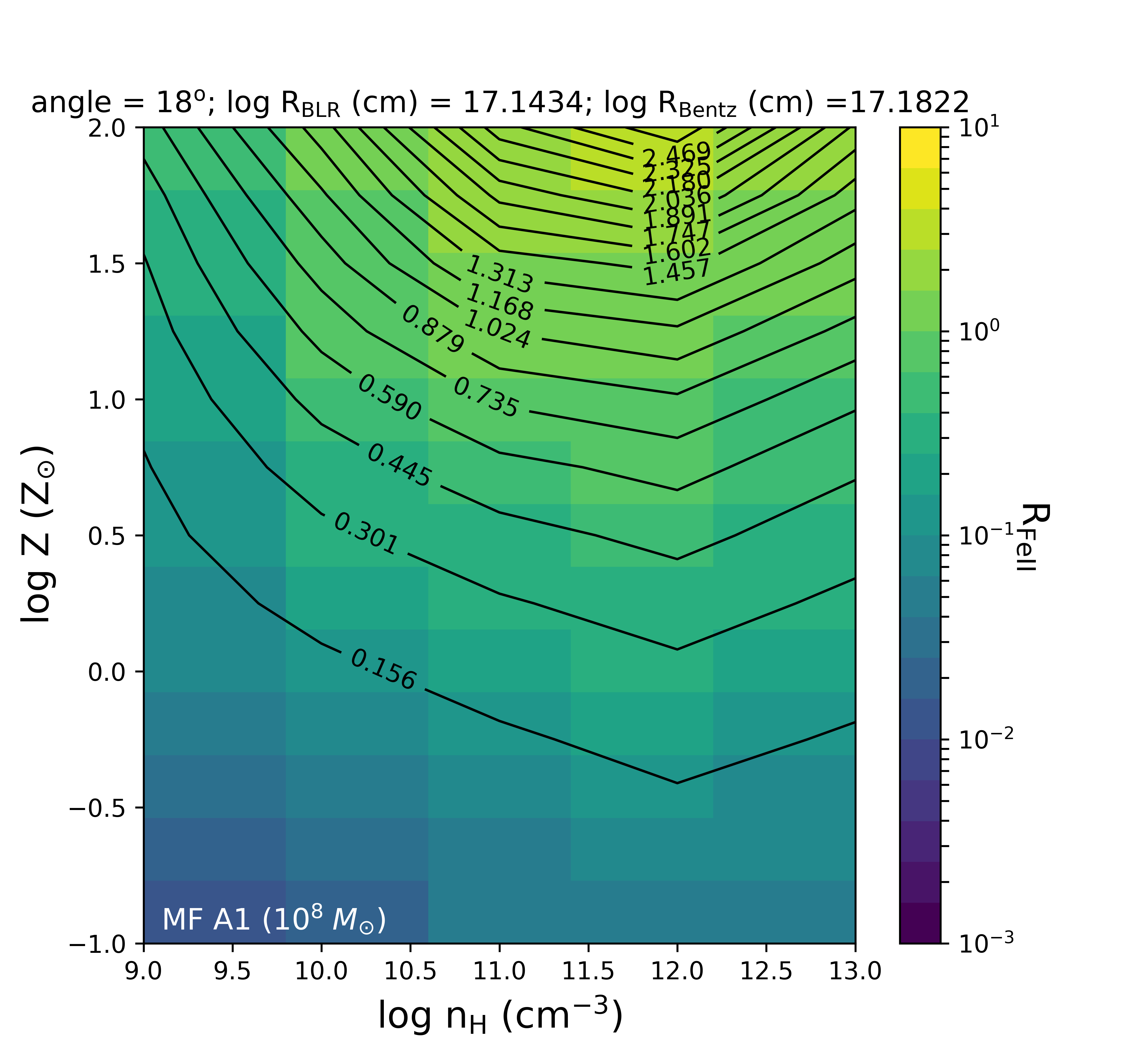}
\end{subfigure}%
\begin{subfigure}{.5\textwidth}
    \centering
    \includegraphics[width=1.075\textwidth]{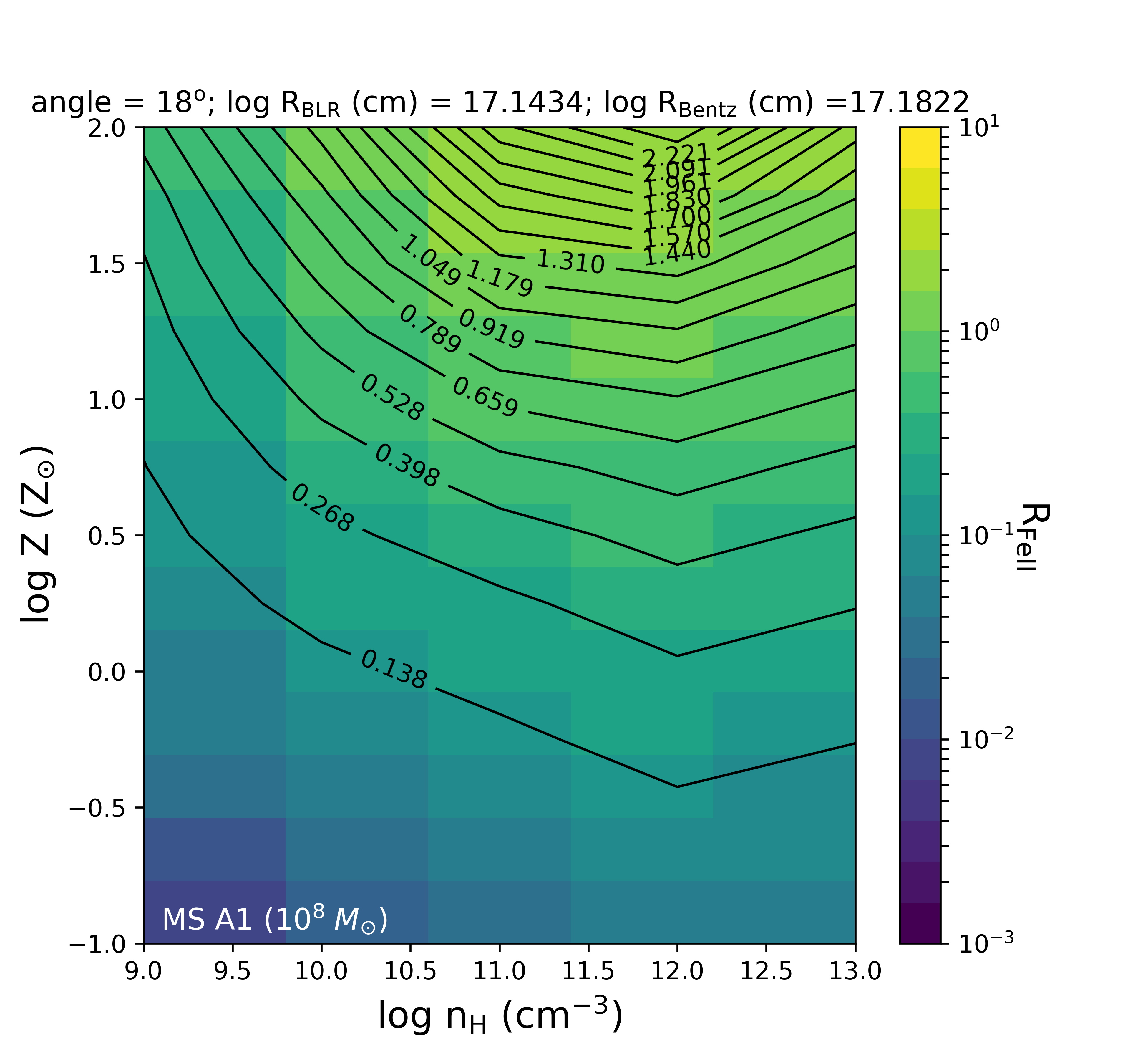}
\end{subfigure}
\caption[short]{Comparison between the four SEDs \citep{kor97, laor97, mf87, ms14} used in the modelling. The figure shows four 2D density plots which map the distribution of the cloud density as a function of the metallicity for a black hole mass $10^8\; M_{\odot}$. The colorbar depicts the value of \rfe\ . This is shown for a representative case (spectral type A1 with FWHM = 2000 \kms) at a viewing angle (here 18 degrees) that corresponds to a inner radius of the BLR which is close agreement to the radius estimation from the standard $r_\mathrm{BLR}$-$L_{\mathrm{5100}}$ relation \citep{bentz13}. The bolometric luminosity is assumed at 0.2$L_\mathrm{Edd}$.}
\label{fig:seds}
\end{figure}

 
 Figure \ref{fig:seds} compares the effect on \rfe\ due to the four different SEDs used in the modelling. It shows four 2D shaded contour plots which map the \rfe\ values  as a function of cloud density and  metallicity for a black hole mass $10^8\; M_{\odot}$. The colorbar depicts the values of \rfe\ . This is shown for a representative case at a viewing angle (18 degrees) that corresponds to a  radius of the BLR which is in close agreement to the radius estimation from the standard $r_\mathrm{BLR}$-$L_{\mathrm{5100}}$ relation \citep{bentz13}. The bolometric luminosity is assumed  0.2$L_\mathrm{Edd}$. The effect of the SEDs we have considered is modest over the full density-metallicity parameter plane. For example, for $n_\mathrm{H} = 10.6$, and $\log Z \approx $0.7 [$Z_\odot$], \rfe\ ranges between 0.4 and 0.6. The largest achievable \rfe\ values are also similar, in the range 2.2 -- 2.5.

\subsection{Effect of microturbulence}

\begin{figure}[b]
\centering
\begin{subfigure}{.5\textwidth}
    \centering
    \includegraphics[width=1.075\textwidth]{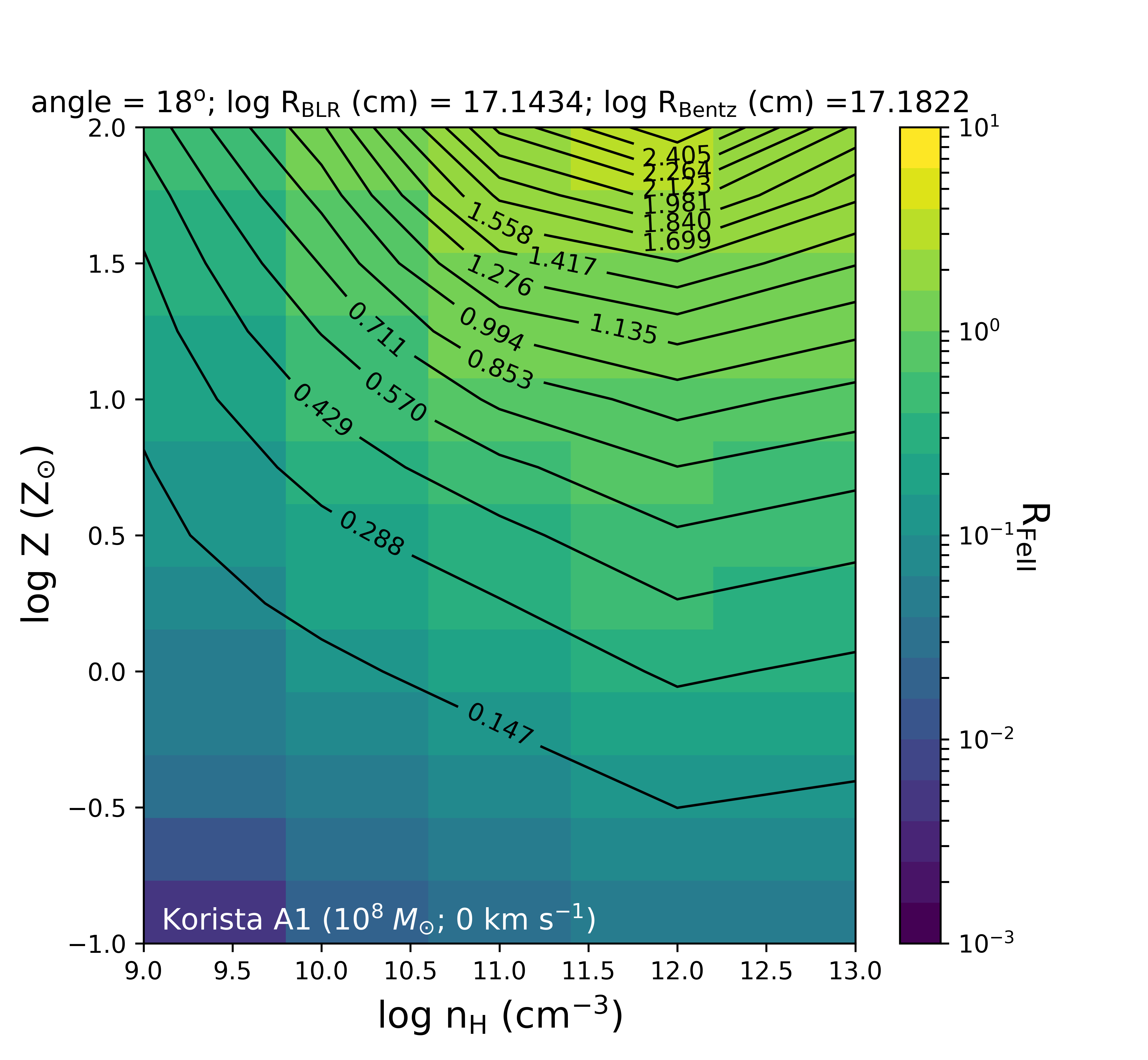}
\end{subfigure}%
\begin{subfigure}{.5\textwidth}
    \centering
    \includegraphics[width=1.075\textwidth]{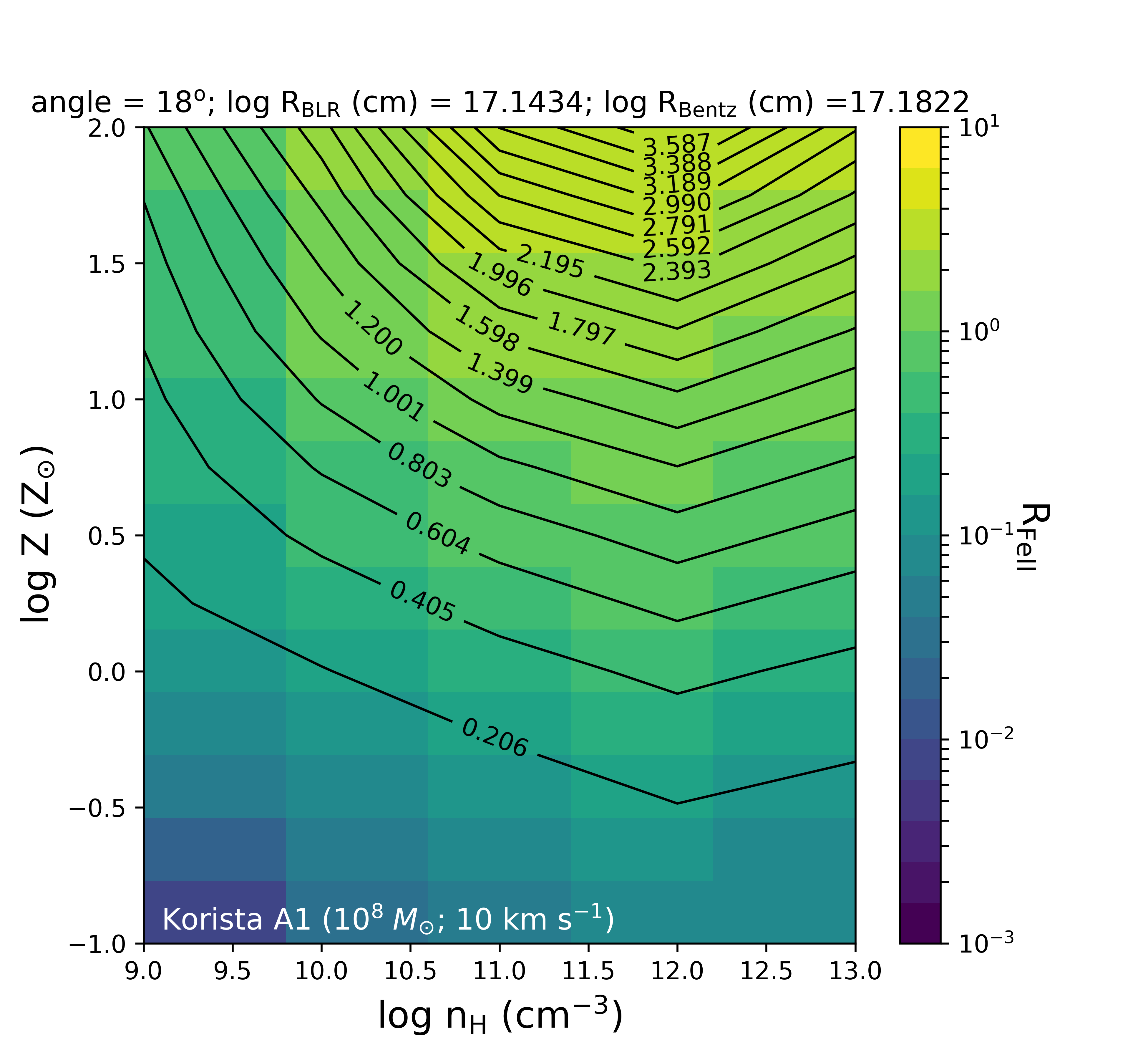}
\end{subfigure}
\vfill
\begin{subfigure}{.5\textwidth}
    \centering
    \includegraphics[width=1.075\textwidth]{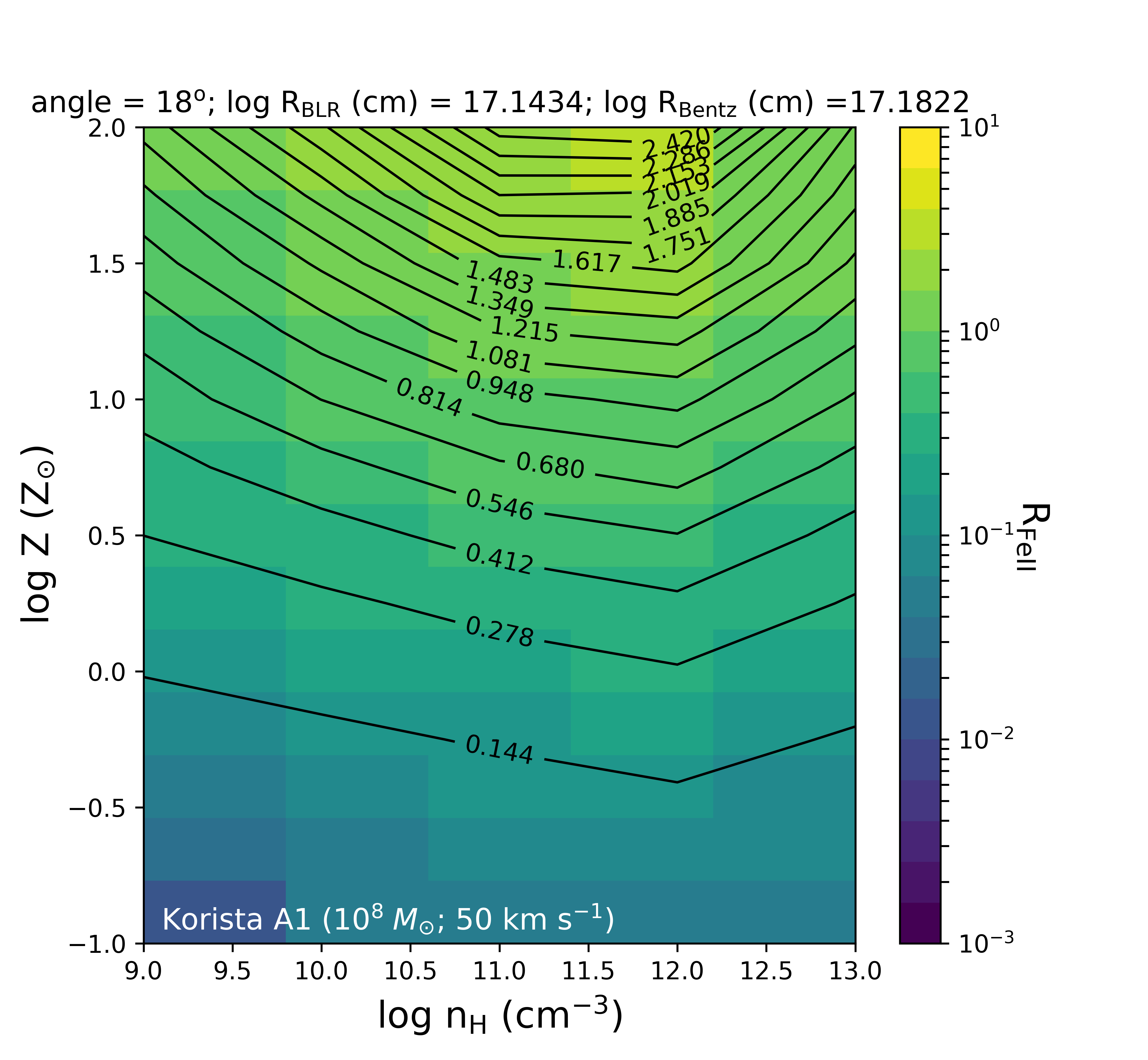}
\end{subfigure}%
\begin{subfigure}{.5\textwidth}
    \centering
    \includegraphics[width=1.075\textwidth]{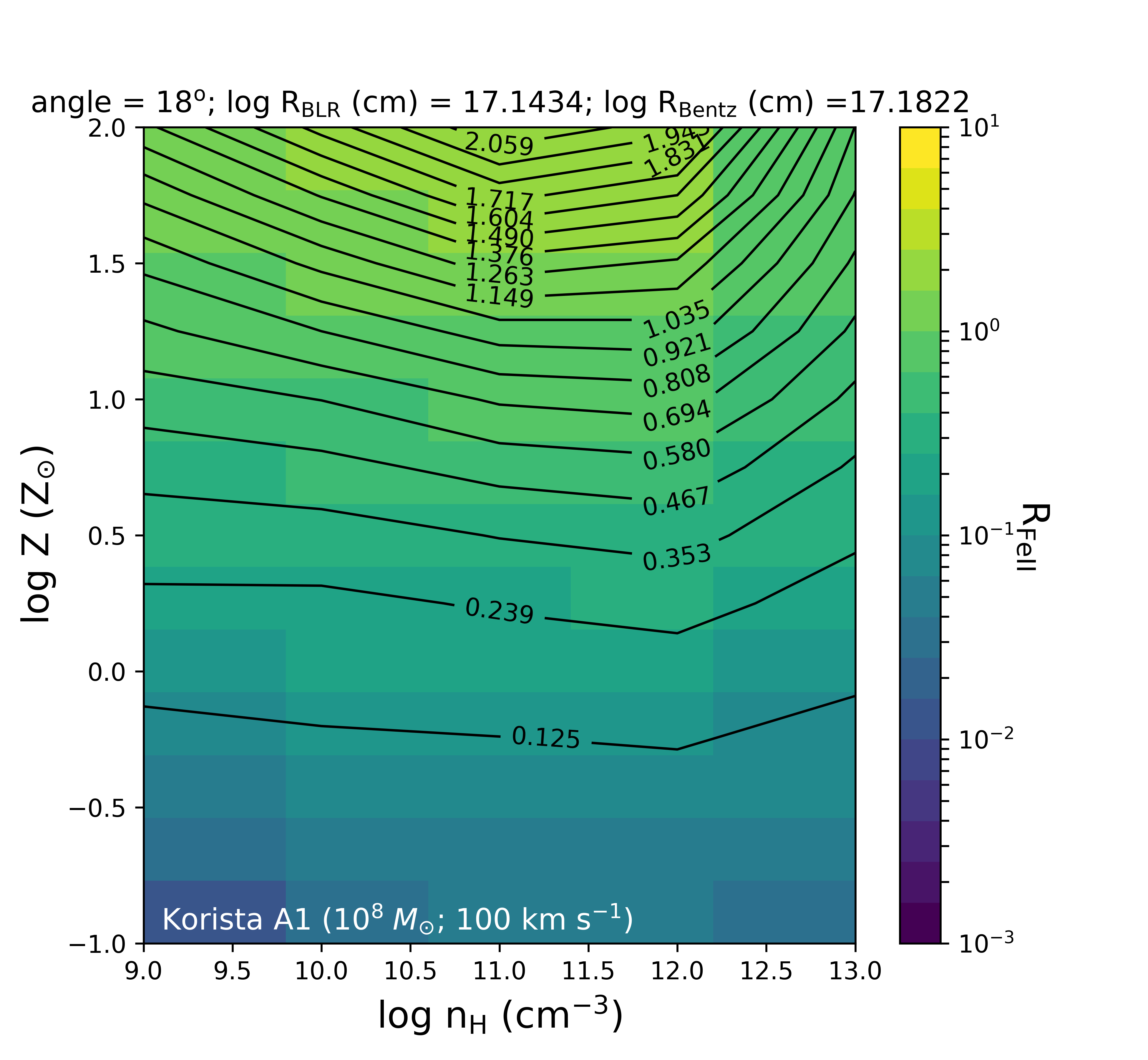}
\end{subfigure}
\caption[short]{Effect of microturbulence -- The figure shows four 2D density plots (for 4 different values of microturbulence: 0, 10, 50 and 100 \kms) which map the distribution of the cloud density as a function of the metallicity for a black hole mass $10^8\; M_{\odot}$. The colorbar depicts the value of \rfe\ . This is shown for a representative case (spectral type A1 with FWHM = 2000 \kms) at a viewing angle (here 18 degrees) that corresponds to a inner radius of the BLR which is close agreement to the radius estimation from the standard $r_\mathrm{BLR}$-$L_{\mathrm{5100}}$ relation \citep{bentz13}. The bolometric luminosity is assumed at 0.2$L_\mathrm{Edd}$ and a representative SED from \citet{kor97} is used.}
\label{fig:turb}
\end{figure}

The effect of micro-turbulence to model the MS has been shown to be of importance \citep{panda18b,panda19}, where the optical plane of quasars is indeed positively affected by inclusion of modest values of micro-turbulence\footnote{there is a $\sim$ 50\% increase in the \rfe\ when the turbulent velocity is increased to 10--20 \kms{}.}. In Figure \ref{fig:turb}, we show the 2D shaded contour plots for 4 different values of microturbulence values, starting from zero turbulent velocity up to 100 \kms. We recover the trends that were first estimated in \citet{panda18b} where we found that the maximum \rfe\ is recovered for the case with a modest value of microturbulence (10--20 \kms). Increasing the microturbulence any higher than these values suppressed the \feii\ emission and thereby resulting in a decrease in the \rfe\ . It was also found that for $v_{turb} = 100$ \kms\ , the \rfe\ reverts to the values that were obtained for the case with zero microturbulence. 

\par 
The effect of the inclusion of microturbulence can be especially important for retrieving the \rfe\ values that correspond to the high accretors - the xA quasars. Also, we have found that there is a intrinsic coupling between the metallicity and microturbulence. In \citet{panda19b}, we have also shown that for theoretical templates that are generated using CLOUDY for the \feii\ pseudo-continuum, there is quite good agreement with the overall line profiles of \feii\ when compared with templates derived from observation (we have currently tested with two sources: an A1 spectral type -- Mrk 335, and a A3 spectral type -- I Zw 1). To retrieve the closest agreement to the spectral line profiles and the corresponding \rfe\ between the observed and theoretical \feii\ templates\footnote{for the comparison between the templates, we have applied a Gaussian broadening of $\sim$800 \kms{}.}, there is a requirement to combine a realistic value of metallicity confirmed from the observations with a certain value of microturbulence. The value of microturbulence varies from case to case and we infer that it is linked with the spectral types. In Panda et al. (in prep.), we will also test the effect of conjugating the increase in mass of the black hole on the MS and the effect that the microturbulence has on it.

\subsection{Effect of increasing $M_\mathrm{BH}$}
Another interesting inference that was drawn from the analyses in \citet{panda19b} was that the higher FWHM sources ($\gtrsim 6000$ \kms) couldn't be explained assuming a fixed mass of $10^8\;M_{\odot}$. The issue becomes relevant for spectral type B1$^{+}$ and above. These sources are with the broadest line profiles and this can be accounted for with increase in the viewing angle although with implausibly large values, $\gtrsim$ 60 degrees. Such large $\theta$\ values are problematic because the MS is made of the Type-1 sources i.e.,  sources that offer an unimpeded view of their central core. There is another way  high FWHM values can be accounted for -- a higher mass of the black hole. We have tested this possibility by considering a case with $M_{BH} = 10^{10}\;M_{\odot}$ alongside the original case with $M_{BH} = 10^{8}\;M_{\odot}$, and comparing the \rfe\ distribution as a function of the viewing angle and correspondingly the radius of the BLR in the vertical bins (see Figure 4 in \citealt{panda19b}). We find that increasing the $M_\mathrm{BH}$ to such masses (consistent for quasars in evolved systems)  increases the net \feii\ emission, and yields viewing angles within the acceptable range for un-obscured sources.

\begin{figure}
\hfill
\vfill
\begin{turn}{90}
\begin{minipage}[c][1\textwidth][c]{1\textheight}
\begin{subfigure}[h]{.475\linewidth}
\includegraphics[width=\linewidth]{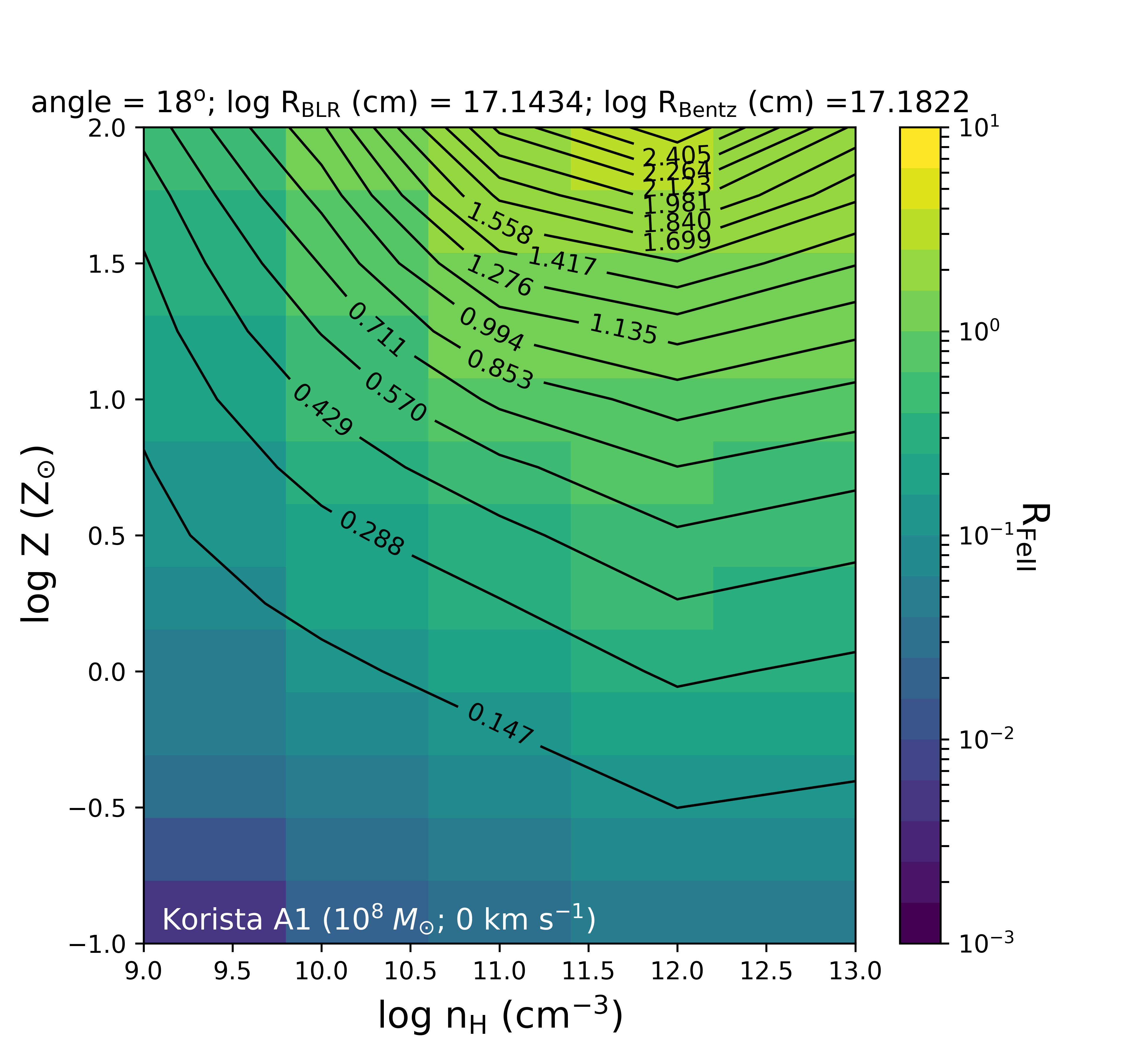}
\end{subfigure}\hfill
\begin{subfigure}[h]{.475\linewidth}
\includegraphics[width=\linewidth]{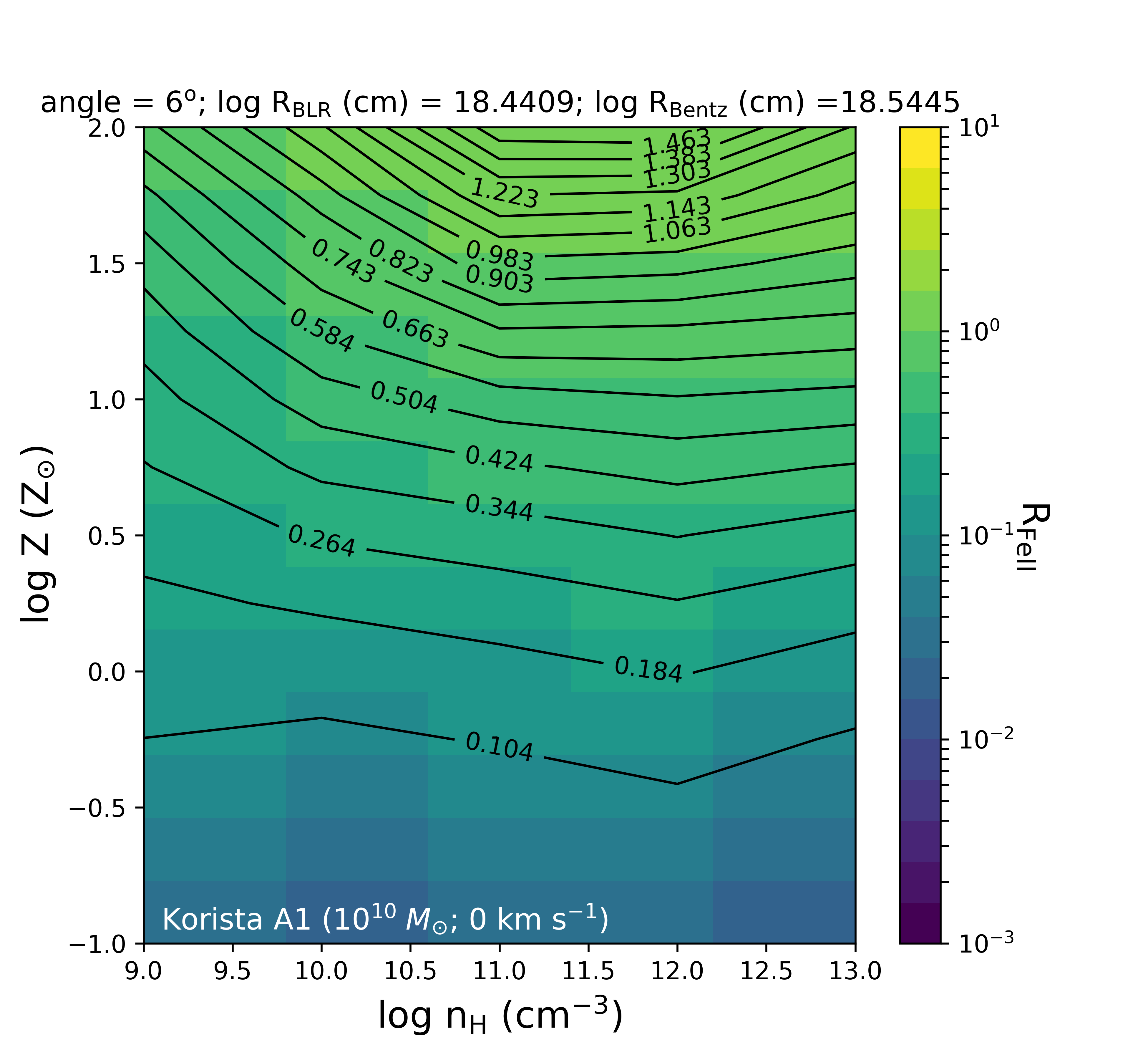}
\end{subfigure}
\caption{Effect of increasing $M_\mathrm{BH}$. The figure shows two 2D density plots which map the distribution of the cloud density as a function of the metallicity. The colorbar depicts the value of \rfe\ . The plots show the results from a set of CLOUDY simulations for two cases of black hole mass ($M_\mathrm{BH})$ (left panel) $10^8\; M_{\odot}$, and (right panel) $10^{10}\; M_{\odot}$. This modelling is performed at zero turbulence, and using an SED shape taken from \citet{kor97}. The plots shown are for a representative case of the spectral type A1 where the mean of the FWHM is assumed at 2000 \kms\ with an assumed Eddington ratio, $\lambda_{\rm{Edd}}$ = 0.2. The corresponding values of the $r_\mathrm{BLR}$ computed from Equation \ref{eq2} and the respective $r_\mathrm{BLR}$ from the standard $r_\mathrm{BLR}$-$L_{\mathrm{5100}}$ relation are shown in the title of each plot.}
\label{fig:montage_m8_m10}
\end{minipage}
\end{turn}
\end{figure}

In Figure \ref{fig:montage_m8_m10}, we show an example of how an increase in the black hole mass (going from $10^8\; M_{\odot}$ to $10^{10}\; M_{\odot}$) affects the parameter space. For this exemplary case, we have assumed a FWHM=2000 \kms{} and Eddington ratio, \lledd{} = 0.2. Yet, this simple test already reveal quite interesting preliminary conclusions. First, due to this increase in the black hole mass, we obtain a much larger $r_\mathrm{BLR}$\ from the standard  \citet{bentz13} relation. To be consistent with our approach, we find the solution for the $r_\mathrm{BLR}$\ from the virial relation that is closest to the $r_\mathrm{BLR}$-$L_{\mathrm{5100}}$ relation, and retrieve back the corresponding value for the viewing angle. We see that the viewing angles in the case of higher black hole mass (here, $10^{10}\; M_{\odot}$) are relatively smaller. From the point of view of the recovered \rfe{} values, there is a $\sim$30-40\% drop, when we go from $10^8\; M_{\odot}$ to $10^{10}\; M_{\odot}$. But, these results are driven by our assumption of fixed FWHM value and the Eddington ratio. In principle, we need an extensive study of the evolution of the parameter space as a function of increasing FWHM and Eddington ratio. This will be reported in Panda et al. (in prep.).

Similar to the interpretation in Section \ref{viewing-angle} and Figure \ref{fig:montage_m8_Kor_A1}, we find that with an increase in the black hole mass under the same parameterization, we see that there is a requirement of higher metallicities for the higher densities ($> 10^{9.75}\; \mathrm{cm^{-3}}$) which goes up by a factor $\sim$2. But, in the lower density regime ($< 10^{9.75}\; \mathrm{cm^{-3}}$), the metallicities required to recover the optimal \rfe{} are lower by factor $\sim$2.5 when going from $10^8\; M_{\odot}$ to $10^{10}\; M_{\odot}$.

\section{Conclusions and the Future: a predictive tool to estimate BLR size?}
\label{conclusions}
We addressed the effect of viewing angle in the accretion disk plane and the geometry of the BLR in the context of the   distribution of quasars in the plane FWHM H$\beta$ -- \rfe. Treating the viewing angle along with a broad range of physically motivated parameters that affect \feii\ emission in Type-1 AGNs i.e., Eddington ratio, local cloud density, metallicity, microturbulence and the shape of the ionizing SED, we have covered the full extent of the quasar main sequence. The values of these physical parameters were known from prior studies.
\par 
In this paper, we summarily described the following results:
\begin{enumerate}
    \item We are now able to constrain the viewing angle for each spectral type corresponding to these sources.
    \item We have incorporated four different SEDs to justify the differences in the ionizing continua observed for a broad distribution of quasars.
    \item The inclusion of the turbulent velocity inside the cloud (microturbulence) recovers the trends that were obtained in \citet{panda18b}: the maximum \feii\ emission and the maximum \rfe\ values  correspond to the case with modest values of microturbulence (10--20 \kms).
    \item We briefly described the effect of increasing black hole mass to explain the sources with high FWHM.
\end{enumerate}

\par 

In \citet{panda19b} and in this paper, we have explored the possibility of constraining the viewing angle for the broad distribution of quasars. Our model can also explain the physical parameters responsible for high accretors which turn out to be predominantly strong \feii{} emitters (the xA sources in Fig. \ref{fig:inset}).  With this model, we can test for reliability by comparing the results with real sources\footnote{We checked this for two sources -- for Mrk 335: 0-10 \kms\ ; I Zw 1: 40-50 \kms\ with super-solar metallicities 1-2.5 Z$_{\odot}$ and 4.8-5.4 Z$_{\odot}$, respectively (see \citealt{panda19b})}. The results we have shown here are a `snapshot' and more detailed analysis will be shown in a forthcoming work (Panda et al. in prep.).

The parameter \rfe\ which can be estimated from a single epoch spectrum for a given source and the metallicity from the line diagnostics from emission lines mostly in the UV \citep{hamann2002,marziani2019}. These data  make it possible to extend our analysis. Values of \rfe\ and metallicity can   be projected in the parameter space maps (see Figure \ref{fig:montage_m8_Kor_A1}) to ultimately recover the virial radius of the broad-line region. Although this possibility needs robust testing,  it {  might be} applicable as a predictor for future reverberation mapping studies. The prediction would be specially valuable for high accretors that show high \rfe ($\gtrsim 1$)  shorter time delays with respect to the ones derived from the scaling law of \citet{bentz13}.  By the same token, we may become able to constrain the $r_\mathrm{BLR}$-$L_{\mathrm{5100}}$ relation that has been shown to offer prospects of application for cosmology by building the Hubble diagram for quasars \citep{lusso15, martinez-aldama19}.

\section*{Acknowledgements}
The project was partially supported by the Polish Funding Agency National Science Centre, project 2017/26/\-A/ST9/\-00756 (MAESTRO  9) and MNiSW grant DIR/WK/2018/12. PM acknowledges the Programa de Estancias de Investigaci\'on (PREI) No. DGAP/\-DFA/\-2192/\-2018 of UNAM, and funding from the INAF PRIN-\-SKA 2017 program 1.05.01.88.\-04. SP and PM would like to acknowledge the organizers of \href{https://matteobachetti.github.io/supereddington2018//}{``Breaking the limits 2018: Super-Eddington accretion onto compact objects''} where this project was first realized. SP would like to acknowledge Mary Loli Mart\'inez Aldama, Deepika Bollimpalli and Abbas Askar for fruitful discussions that helped in the realization of the project. We thank the anonymous referee for useful comments that helped to improve the content of the paper.

\bibliography{references}

\end{document}